\documentstyle[12pt]{article}
\newfont{\g}{eufm10 scaled\magstep1}

\newcommand{\gtS}{\mbox{\g S}}
\newcommand{\gtgo}{\mbox{\g go}}
\newcommand{\gto}{\mbox{\g o}}
\newcommand{\sukima}{\mbox{\hspace{4mm}}}
\newcommand{\nequiv}{\equiv \!\!\!\!\! /\mbox{\hspace{1mm}}}

\def\p{\partial}
\def\l{\lambda}
\def\L{\Lambda}
\def\d{\delta}

\def\fock{{\cal F}}
\def\bra{\langle}
\def\ket{\rangle}
\def\A{$A^{(2)}_{2l}$}
\def\D{$D^{(2)}_{l+1}$}

\begin{document}
\thispagestyle{empty}
\rightline{TMUP-HEL-9403}
\rightline{January, 1994}
\vspace{1cm}
\begin{Large}
\begin{center}
Basic Representations of \A and \D and

the Polynomial Solutions to the Reduced BKP Hierarchies
\end{center}
\end{Large}

\vspace{1.0cm}

\begin{center}

Tatsuhiro NAKAJIMA

{\it Department of Physics, Tokyo Metropolitan University}

{\it Minami-Ohsawa 1-1, Hachioji, Tokyo 192-03, Japan}

and

Hirofumi YAMADA

{\it Department of Mathematics, Tokyo Metropolitan University}

{\it Minami-Ohsawa 1-1, Hachioji, Tokyo 192-03, Japan}

\end{center}

\vspace{1.0cm}

\begin{center}
{\bf Abstract}
\end{center}

Basic representations of \A and \D are studied.
The weight vectors are represented in terms of
Schur's $Q$-functions. The method to get the polynomial solutions
to the reduced BKP hierarchies is shown to be equivalent to
a certain rule in Maya game.
\newpage
\begin{flushleft}
{\large\bf \S 1 \hspace{5mm} Introduction}
\end{flushleft}
\vspace{5mm}

Nonlinear integrable systems are important in mathematical physics.
Among others the KP hierarchy
has been deeply studied in various aspects such as representation
theory, algebraic geometry and two dimensional gravity.
In 1981 Date {\it et al.} introduced a KP like hierarchy of nonlinear
differential equations, which has a symmetry of the infinite
dimensional Lie algebra
$B_{\infty}=\gtgo(\infty)$ and call it the KP hierarchy of
$B$ type or BKP hierarchy for short \cite{DKM,DJKM1}.
Although this hierarchy has nice properties which reflect
the Lie algebra symmetry, it seems that a few researches have been made
compared with the KP hierarchy. This letter is an attempt to
give an explicit expression of the weighted homogeneous solutions
to the reduced BKP hierarchies.

We first review some ingredients of the
BKP hierarchy and affine Lie algebras for
the following discussions.

Let $\phi_n$ ($n \in {\bf Z}$) be the neutral free fermion satisfying
\begin{equation}
\{\phi_m,\phi_n\}=(-1)^m \delta_{m+n,0}.
\end{equation}
Note that $\phi_0^2=1/2$.
We denote by ${\bf B}={\bf B}_0\oplus{\bf B}_1$ the superalgebra
generated by the neutral free fermions.
Let $\fock=\fock_0\oplus \fock_1={\bf B}_0|0\ket\oplus{\bf B}_1|0\ket$
(resp. $\fock^* = \bra0|{\bf B}_0\oplus\bra0|{\bf B}_1$) be the Fock space
(resp. dual Fock space), where the vacuum $|0\ket$ (resp. $\bra0|$)
is defined by
\begin{equation}
\phi_n|0\ket=0 \mbox{\hspace{4mm}}(n<0),\mbox{\hspace{5mm}}
\mbox{(resp.\hspace{2mm}}\bra0|\phi_n=0 \mbox{\hspace{4mm}}(n>0)).
\end{equation}
The vacuum expectation value $\bra0|a|0\ket$ ($a\in {\bf B}$)
is uniquely determined by setting
$\bra0|1|0\ket=1$, $\bra0|\phi_0|0\ket=0$.

We construct the realization of $\fock$.
The normal ordering for the neutral fermions is defined by
\begin{equation}
:\phi_i \phi_j:=\phi_i \phi_j-\bra0|\phi_i \phi_j|0\ket.
\end{equation}
Define the Hamiltonian by
\begin{equation}
H(t)={1\over2}\sum_{j\geq1,odd} \sum_{n\in{\bf Z}}
(-1)^{n+1} t_j \phi_n \phi_{-j-n}.
\end{equation}
Let $V={\bf C}[t_1,t_3,t_5,\ldots]$ be a polynomial ring.
There is an isomorphism between $\fock_0$ and $V$, defined by
\begin{equation}
a|0\ket \longmapsto \bra0|e^{H(t)}a|0\ket,
\mbox{\hspace{3mm}}a\in {\bf B_0}.
\end{equation}
The subspace
\begin{equation}
B_{\infty}=\left\{\sum_{m,n\in{\bf Z}} c_{mn}:\phi_m \phi_n:
; c_{mn}=0 \mbox{ if } |m-n|>>0 \right\}
\end{equation}
of ${\bf B}_0$ admits a structure of a Lie algebra isomorphic to the
one dimensional central extension of $\gto(\infty)$.
The representation of $B_{\infty}$ on $V$ is defined
by the vertex operator
\begin{equation}
Z(p,q)={1\over2}\frac{1-q/p}{1+q/p}
\left\{\exp\left(\sum_{j:odd} (p^j-q^j) t_j \right)
\exp\left(-2 \sum_{j:odd}{1\over j} (p^{-j}-q^{-j}) {\p\over{\p t_j}}
 \right)-1\right\},
\label{eqn:vertex}
\end{equation}
which corresponds to the action of $:\phi(p)\phi(q):$ on $\fock_0$, where
$\phi(p)=\sum_{n \in {\bf Z}} \phi_n p^n$.
The space of the solutions ($\tau$-functions) to the BKP hierarchy is
obtained as the group orbit through $1\in V$ under this representation.

We consider the reduction of the BKP hierarchy \cite{DJKM2}. For a fixed
positive integer $r$, $r$-reduced BKP hierarchy
is defined through the specialization of the vertex operator
(\ref{eqn:vertex})
\begin{equation}
q=p\omega, \mbox{\hspace{4mm}} \omega^r=1, \omega\neq1.
\end{equation}
This specialization is known to be equivalent to restricting
the algebra $B_{\infty}$ to
the subalgebra \A for $r=2l+1$,
and \D
for $r=2l+2$. The representation space will be denoted by $V_r$.
It should be noted that the situation is different between
odd $r$ and even $r$. In the $r=2l+1$ case
the specialization means the deletion of the variables
$t_j$($j\equiv0$ mod $r$).
Hence the representation space becomes smaller:
$V_r={\bf C}[t_j; j\geq 1, \mbox{odd and }j\nequiv0 $(mod $r)]$.
On the other hand, in the $r=2l+2$ case, since the representation
space has no variables with an even index,
the space itself is unchanged: $V_r=V$.

In order to compute the weight of a given vector,
we write down coroots of \A and \D explicitly
by means of the neutral fermion operators:
\begin{eqnarray*}
\lefteqn{A^{(2)}_{2l}\mbox{\hspace{3mm}} (r=2l+1)} \nonumber \\
& &\alpha_0^{\vee}=2\sum_{n\in {\bf Z}}
(-1)^{n+1}:\phi_{nr-1}\phi_{-nr+1}:+1,\\
& &\alpha_i^{\vee}=
\sum_{n\in{\bf Z}}(-1)^{n+i}\left(:\phi_{nr+i}\phi_{-nr-i}:-
:\phi_{nr-(i+1)}\phi_{-nr+(i+1)}: \right) (1\leq i \leq l-1),\\
& &\alpha_l^{\vee}=\sum_{n\in{\bf Z}}(-1)^{n+l}:\phi_{nr+l}\phi_{-nr-l}:.\\
\lefteqn{D^{(2)}_{l+1} \mbox{\hspace{3mm}}(r=2l+2)}\nonumber \\
& &\alpha_0^{\vee}=-2\sum_{n\in {\bf Z}}:\phi_{nr-1}\phi_{-nr+1}:+1,\\
& &\alpha_i^{\vee}=\sum_{n\in{\bf Z}}(-1)^i\left(:\phi_{nr+i}\phi_{-nr-i}:-
:\phi_{nr-(i+1)}\phi_{-nr+(i+1)}: \right) (1\leq i \leq l-1),\\
& &\alpha_l^{\vee}=2\sum_{n\in{\bf Z}}(-1)^l:\phi_{nr+l}\phi_{-nr-l}:.
\end{eqnarray*}

Let $\delta=\sum_{i=0}^l a_i \alpha_i$ be the fundamental imaginary root
of the affine Lie algebra, where $a_i$ are the labels of the
corresponding generalized Cartan matrix.
For \A and \D it reads $\delta=2\sum_{i=0}^{l-1}
\alpha_i + \alpha_l$ and $\delta=\sum_{i=0}^{l}\alpha_i$, respectively.

Finally we recall the basic representation of affine Lie algebras.
The irreducible representation with highest weight $\L_0$ is called the
basic representation, where
\begin{equation}
\L_0(\alpha_0^{\vee})=1, \mbox{\hspace{4mm}}\L_0(\alpha_j^{\vee})=0
\mbox{\hspace{2mm}}(j\neq0).
\end{equation}
The representation on $\fock_0$ constructed above turns out
to be the basic representation $L(\L_0)$ of \A and \D \cite{JM}.
\vspace{5mm}

\begin{flushleft}
{\large\bf \S 2 \hspace{5mm} Weight Vectors, Q-Functions and Maya Game}
\end{flushleft}
\vspace{5mm}

It is shown in \cite{JM} that a vector
\begin{equation}
(m_1,\ldots m_{2d}):=
\phi_{m_{2d}}\cdots \phi_{m_1}|0\ket \in \fock_0,
\sukima (m_{2d}>\cdots>m_1 \geq0)
\label{eqn:vec}
\end{equation}
is a weight vector of the Fock representation of $B_{\infty}$. Since
\A and \D are Lie subalgebras of $B_{\infty}$, the vector (\ref{eqn:vec})
is also a weight vector of $L(\L_0)$ of these affine Lie algebras. In view
of the principally specialized character of $L(\L_0)$,
we have the following proposition.

\noindent{\bf Proposition 1}
{\it The basic representation $L(\L_0)$ of \A (resp. \D)
has the basis consisting of the weight vectors}
\begin{eqnarray}
& &\left\{(m_1,\ldots, m_{2d}); m_{2d}>\cdots>m_1\geq0,
m_j\nequiv0 (\mbox{mod} 2l+1) \right\} \\
& &\left(\mbox{\it resp. }
\left\{(m_1,\ldots,m_{2d}); m_{2d}>\cdots>m_1\geq0 \right\}
\right).
\end{eqnarray}

Let us turn to the realization $V_r$ of the basic representation
$L(\L_0)$, where $r=2l+1$ for \A and $r=2l+2$ for \D. We see how each
weight vector of $L(\L_0)$ is expressed explicitly as a polynomial in $V_r$.
To this end we recall Schur's $Q$-functions \cite{Mac}.
Let $X_1,\ldots,X_n$ be indeterminates. The Hall-Littlewood symmetric
function indexed by the Young diagram $Y=(m_1,\ldots,m_n)$
$(m_n\geq \cdots \geq m_1\geq 0)$ is defined by
\begin{equation}
Q_Y(X_1,\ldots,X_n; q)=\prod_{j\geq1}(q; q)_{\mu_j}
\sum_{w\in \gtS_n/\gtS_n^Y}w\left(X_1^{m_n}\cdots X_n^{m_1}
\prod_{m_i<m_j}\frac{X_i-q X_j}{X_i-X_j} \right)
\label{eqn:defq}
\end{equation}
where $q$ is a parameter, $\mu_j=\#\{i; m_i=j\}$ is the multiplicity
of $j$ in $Y$,
$(q;q)_k=\prod_{i=1}^{k}(1-q^i)$ and $\gtS_n^Y$ is the subgroup consisting
of permutation $w$ which does not change the diagram $Y$,
{\it i.e.}, $m_{w(i)}=m_i$ for $1\leq i\leq n$.
As a specialization $q=-1$ we get Schur's $Q$-function $Q_Y(X_1,\ldots,X_n)$.
It is obvious from (\ref{eqn:defq}) that $Q_Y(X_1,\ldots,X_n)=0$
unless $Y$ is a strict Young
diagram, {\it i.e.}, $\mu_j=1$ for any $j\geq 1$.

The Frobenius formula gives the relation between $Q$-functions
and power sum symmetric functions
$p_m(X_1,\ldots,X_n)=X_1^m+\cdots+X_n^m$:
\begin{equation}
Q_Y(X_1,\ldots,X_n)=\sum_{Y'} 2^{l(Y')} z_{Y'}^{-1} \chi^Y_{Y'}(-1)
\prod_{j=1}^n p_{m_j'}(X_1,\ldots,X_n).
\end{equation}
Here the summation runs over all Young diagrams $Y'=(m_1',\ldots,m_n')$
consisting of odd numbers $m_j'$, $l(Y')$ is the number of
nonzero $m_j'$'s, and $z_{Y'}=\prod_{j\geq1} j^{\nu_j} \nu_j!$,
where $\nu_j$ denotes the multiplicity of $j$ in $Y'$. Although
we do not make explicit, $\chi^Y_{Y'}(-1)$ is an integer computed
easily from the character table of projective representations of
the symmetric group $\gtS_n$. Introduce new variables
$t_j=\frac{2}{j}p_j(X_1,\ldots,X_n)$ for positive odd integers $j$,
so that the $Q$-function turns out to be a weighted homogeneous polynomial
of $t=(t_1, t_3, \ldots)$:
\begin{equation}
Q_Y(t)=\sum_{Y'}\chi^Y_{Y'}(-1)
\frac{t_1^{\nu_1}t_3^{\nu_3}\cdots}{\nu_1!\nu_3!\cdots} \in V.
\end{equation}

Another expression is known for the $Q$-functions \cite{HH}.
Let $Y=(m_1,\ldots,m_{2d})$ ($m_{2d}>\cdots>m_1\geq 0$)
be a strict Young diagram, and consider
$2d\times 2d$ skew symmetric matrix $\left(Q_{(m_i,m_j)}(t)\right)$.
Then we have
\begin{equation}
Q_Y(t)=\mbox{Pf}\left(Q_{(m_i,m_j)}(t)\right),
\end{equation}
where Pf denotes the Pfaffian.
Using anti-commutation relations, it is easily deduced that
\begin{equation}
\bra0|e^{H(t)}\phi_m\phi_n|0\ket={1\over2}Q_{(m,n)}(t) \sukima (m>n\geq0)
\end{equation}
and
\begin{equation}
\bra0|e^{H(t)}\phi_{m_{2d}}\cdots\phi_{m_1}|0\ket=
{1\over{2^d}}\mbox{Pf}\left(Q_{(m_i,m_j)}(t) \right).
\end{equation}
Hence every weight vector of the Fock representation of $B_{\infty}$
is expressed by means of a single $Q$-function.

For a polynomial $P(t)\in V$, we denote by $P(t')$ the polynomial
in $V_r$ obtained from $P(t)$ by putting $t_{jr}=0$ for $j=1, 2, \ldots$.
Our claim is as follows:

\noindent{\bf Proposition 2}
{\it The basic representation on $V_r$ of \A with $r=2l+1$
(resp. \D with $r=2l+2$) has the basis consisting of weight vectors}
\begin{eqnarray}
& &\left\{Q_Y(t'); Y=(m_1,\ldots,m_{2d}), m_{2d}>\cdots>m_1\geq0,
m_j \nequiv 0 (\mbox{mod} 2l+1) \right\} \\
& &\left(\mbox{resp. }
\left\{Q_Y(t'); Y=(m_1,\ldots,m_{2d}), m_{2d}>\cdots>m_1\geq0 \right\}\right).
\end{eqnarray}

Maya game is a fermion version of the so called ``Nim'' defined as follows
\cite{ichimatsu}.
Consider infinitely many cells indexed by ${\bf Z}$. Each $m\in{\bf Z}$
is said to be either black or white according as it is filled with a single
particle or is empty.
The boundary condition is  that $m$(resp. $-m$) is white(resp. black)
for sufficiently large $m>0$. The vacuum is a state that $m$ is black
for all $m<0$ and is white for all $m\geq0$. (Figure 1)
In general state some particles
excite and occupy some non-negative numbered cells.
The game, played by two, is to move a particle by turn to a lower level until
one completes the vacuum. The player loses the game if he faces the
stalemate, {\it i.e.}, the  vacuum state.

To each weight vector $(m_1,\ldots,m_{2d})$
we associate a state of Maya game in a following manner.
Every positive cell $m_j$
is filled with a particle excited from the cell $-m_j$. There is an ambiguity
at the cell indexed $0$. We choose the convention that $0$ is
always white. The state of Maya game such obtained from a weight vector
is called a symmetric state.

In the language of Young diagrams the relation between symmetric
states and weight vectors is described as follows. For a weight vector
$(m_1,\ldots,m_{2d})$, draw the corresponding strict Young diagram $Y$.
Let $\widetilde{Y}$ be the shift symmetric Young diagram
obtained from $Y$, which is explained in \cite{Mac}. Then the rule
that associates a state of Maya game with a Young diagram leads to
the symmetric state corresponding to $(m_1,\ldots,m_{2d})$
(cf. \cite{ichimatsu}). (Figure 2)
We remark that the transposed Young diagram $^t\widetilde{Y}$ leads to
the symmetric state with black $0$.

We now state how to obtain weight vectors of weight $\l\pm \delta$
from a given weight vector of weight $\l$. The procedure is  described
by modifying the rule of Maya game.

First we consider the case of \A. Suppose a weight vector of weight
$\l$ is given. Then one of the following possible moves is
permitted to have weight $\l-\delta$.

(1) Move a black by $2l+1$ to the right.

(2) Replace the pair of white cells $(k, 2l+1-k)$ by black for
$1\leq k \leq l$.

\noindent
For the case \D the moves are

(1) Replace the white $l+1$ by black.

(2) Move a black at a positive multiple of $l+1$ to the right.

(3) Once the cell $n(2l+2)$ becomes black for an $n\geq1$, instead of
that single black, the black pairs $(k,n(2l+2)-k)$ are also permitted for
$1\leq k<n(2l+2)$.

The inverse moves give weight vectors of weight $\l+\delta$ from
one of weight $\l$.

\begin{flushleft}
{\large\bf \S 3 \hspace{5mm} Polynomial solutions
to the reduced BKP hierarchies}
\end{flushleft}
\vspace{5mm}

In this section we characterize the maximal weight vectors
of the basic representation $L(\Lambda_0)$ of the affine Lie
algebras \A and \D and, as a consequence,
obtain the weighted homogeneous polynomial solutions to the
corresponding reduced BKP hierarchy.

By definition the weight $\l$ of $L(\Lambda_0)$ of \A
or \D is said to be maximal if $\l+\delta$ is not a
weight. It is known that the totality of the maximal weights of
$L(\Lambda_0)$ is given by the Weyl group orbit through the highest
weight $\Lambda_0$, and hence each maximal weight is of multiplicity $1$
\cite{Kac}. Since the Weyl group can be seen as a
subgroup of the group corresponding to the affine Lie algebra,
the maximal weight vectors
of $L(\Lambda_0)$ of \A and \D are polynomial
solutions to the $(2l+1)$-reduced and $(2l+2)$-reduced BKP hierarchy,
respectively \cite{DJKM2}.

As we have shown in section 2, for each weight vector in $L(\L_0)$
we can associate a symmetric state of Maya game. We have also given
a description by means of Maya game how to obtain weight vectors of
weight $\l \pm \d$ for a given weight vector of weight $\l$. The weight $\l$
is maximal if the corresponding weight vectors is at stalemate
according to the rule.

Let us give a more precise description. For \A the weight
vector $(m_1,\ldots,m_{2d})$ is a maximal weight vector if each $m_j$
cannot be moved by $2l+1$ to the left and there are no such pairs
$(m_i,m_j)=(k,2l+1-k)$ for $1\leq k \leq l$.
For \D, $(m_1,\ldots,m_{2d})$ is a maximal weight vector if
there are no $m_j$ such that $m_j\equiv 0$ (mod $l+1$) and there are no pairs
$(m_i,m_j)$ such that $m_i+m_j=n(2l+2)$ for $n\geq 1$.
In particular, for the case
$A^{(2)}_2$ the following set covers the maximal weight vectors:
\[
\{\phi,(1,4,7,\ldots,3n-2), (2,5,8,\ldots,3n-1) ; n \geq 1\}.
\]
For $D^{(2)}_4$ the maximal weight vectors of degree up to $12$ are
\begin{eqnarray*}
& &\{\phi,(1),(2),(3),(1,2),(1,3),(5),(2,3),(6),(1,5),(1,2,3),(7),(1,6),\\
 & &(2,5),(1,2,5),(2,7),(3,6),(1,9),(3,7),(1,3,6),(5,6),(2,10),(5,7),\\
 & &(1,2,9),(1,5,6),(2,3,7)\}.
\end{eqnarray*}
If one uses the language of Young diagrams, the above criterion is
stated in simpler words.

\noindent{\bf Theorem}
{\it Schur's $Q$-function $Q_Y(t')$ is a maximal weight vector of $L(\L_0)$
of \A (resp. \D) if the shift symmetric
diagram $\widetilde{Y}$ of $Y$ has no hooks of length divisible by $2l+1$
(resp. $2l+2$).}

In both cases, \A and \D, one recovers the
(non-reduced) BKP hierarchy by taking the limit $l\rightarrow \infty$.
Hence, as a corollary to the above theorem, we see that the $Q$-function
$Q_Y(t)$ of any strict Young diagram $Y$ solves the BKP hierarchy \cite{You}.

\newpage
\newcounter{num}
\setcounter{num}{-8}
\unitlength 10mm
\begin{picture}(16,4)
\put(0,0){\line(1,0){16}}
\put(0,1){\line(1,0){16}}
\multiput(0.5,0)(1,0){16}{\line(0,1){1}}
\multiput(1,0.5)(1,0){7}{\circle*{0.7}}
\multiput(0.8,1.3)(1,0){15}{
\addtocounter{num}{1}%
\makebox(0,0){\large{\arabic{num}}}}
\end{picture}

\vspace{1cm}
\noindent
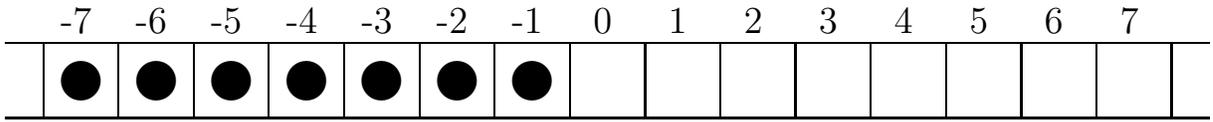
{\large {\bf Figure 1.} Maya diagram of the vacuum state.}

\vspace{2cm}

\setcounter{num}{-8}
\begin{picture}(17,2)
\put(0,0){\line(1,0){16}}
\put(0,1){\line(1,0){16}}
\put(9,0.5){\circle*{0.7}}
\put(12,0.5){\circle*{0.7}}
\multiput(0.5,0)(1,0){16}{\line(0,1){1}}
\multiput(1,0.5)(1,0){3}{\circle*{0.7}}
\multiput(5,0.5)(1,0){2}{\circle*{0.7}}
\multiput(0.8,1.3)(1,0){15}{
\addtocounter{num}{1}%
\makebox(0,0){\large{\arabic{num}}}}
\end{picture}

\begin{center}
{\Large $\Updownarrow$}
\end{center}

\begin{picture}(7,8)
\put(4,7){\large {\boldmath  $Y=$}}
\put(4,3){\large {\boldmath  $\widetilde{Y}=$}}
\put(6,8){\line(1,0){4}}
\put(6,7){\line(1,0){4}}
\put(6,6){\line(1,0){1}}
\multiput(6,7)(1,0){5}{\line(0,1){1}}
\multiput(6,6)(1,0){2}{\line(0,1){1}}
\put(6,5){\line(1,0){5}}
\put(6,4){\line(1,0){5}}
\put(6,3){\line(1,0){3}}
\put(6,2){\line(1,0){1}}
\put(6,1){\line(1,0){1}}
\multiput(6,1)(1,0){2}{\line(0,1){4}}
\multiput(8,3)(1,0){2}{\line(0,1){2}}
\multiput(10,4)(1,0){2}{\line(0,1){1}}
\end{picture}

\noindent
{\large{\bf Figure 2.}
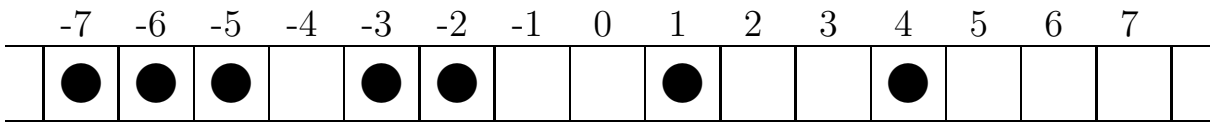
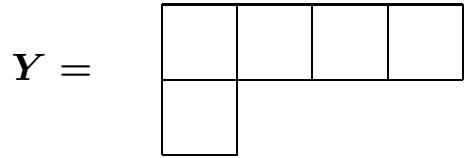
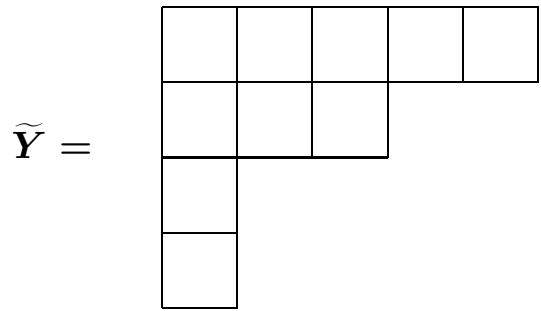
The symmetric state for $(1,4)$ and the corresponding
Young diagrams.}
\end{document}